\documentclass[aps,prl,twocolumn,showpacs,superscriptaddress]{revtex4-2}
\usepackage{latexsym}
\usepackage{amssymb}
\usepackage{graphicx}
\usepackage{amsmath}
\usepackage{bm}
\usepackage[colorlinks,
          linkcolor=black,
            citecolor=black,
            urlcolor=blue
           ]{hyperref}
\usepackage{verbatim}
\usepackage{mathrsfs}
\usepackage{extarrows}
\usepackage{comment}
\usepackage{mathtools,slashed}

\usepackage{cancel}
\usepackage{subfigure}
\usepackage{graphicx}
\usepackage{caption}
\usepackage{ragged2e}

\newcommand{\R}{R_\pi^x}

\begin{document}

\title{Identification of gapless phases by a twisting operator}

\author{Hang Su}
\affiliation{School of Physics and Astronomy, Shanghai Jiao Tong University, Shanghai 200240, China}
\author{Tengzhou Zhang}
\affiliation{School of Physics and Astronomy, Shanghai Jiao Tong University, Shanghai 200240, China}
\author{Yuan Yao}
\email{smartyao@sjtu.edu.cn}
\affiliation{School of Physics and Astronomy, Shanghai Jiao Tong University, Shanghai 200240, China}
\author{Akira Furusaki}
\affiliation{Quantum Matter Theory Research Group, RIKEN CEMS, Wako, Saitama 351-0198, Japan}

\begin{abstract}
We propose a general necessary condition for a spin chain with SO(3) spin-rotation symmetry to be gapped. 
Specifically, we prove that the ground state(s) of an SO(3)-symmetric gapped spin chain must be spin singlet(s), 
and the expectation value of a twisting operator asymptotically approaches unity in the thermodynamic limit, where finite-size corrections are inversely proportional to the system size. 
This theorem provides (i) supporting evidence for various conjectured gapped phases,
and (ii) a sufficient criterion for identifying gapless spin chains.
We verify our theorem by numerical simulations for a variety of spin models 
and show that it offers a novel efficient way to identify gapless phases in spin chains with spin-rotation symmetry.
\end{abstract}

\date{\today}

%%%%%%%%%%%%%%%%%%%%

\maketitle

\paragraph{Introduction.---}
Identifying and classifying various phases of quantum many-body systems is a subject of great interest in statistical and condensed matter physics~\cite{Landau:1937aa}. 
Characteristic features of those quantum systems are observed in their low-energy sector, %properties
e.g., ground-state wave function(s) and various low-lying excitations,
%However, non-perturbative effects of many-body interactions and strong correlations make the task of phase identification complicated.
and roughly speaking,
the quantum many-body systems are classified into \textit{gapped} and \textit{gapless} phases based on their distinct low-energy spectra~\cite{Zeng:2015aa,zeng2019quantum}.
%and distinguishing them in a phase diagram is a basic step in the classifications of quantum phases.
%===========
%There have been significant progresses in gapped phases,
%e.g., symmetry-protected topological (SPT) phase classifications and topological ordered phases~\cite{Gu:2009aa,Chen:2010aa,Pollmann:2012aa,Wen:TOreview2013}.
%For gapless systems,
%various exotic critical phases and quantum phase transitions have been proposed.

%In parallel, a central theme in the study of quantum phases is the characterization of order. 
%The symmetry breaking is diagonsed through local order parameters~\cite{Landau:1937aa}, 
%while $nonlocal$ probes is needeed to identify SPT phases~\cite{Wen:1990aa,Haegeman:2012aa,Pollmann:2012ab}, 
%like string order parameters through their nonvanishing ground-state expectation values~\cite{Nijs:1989aa,Oshikawa1992,Lou:2003aa,Duivenvoorden:2012aa,Ueda:2014aa,PhysRevB.90.235111,Ogino:2022aa}. 
%More recently, rigorous topological indices have been proposed for one-dimensional gapped spin chains with \textit{unique} gapped ground states~\cite{Nakamura:2002aa,Tasaki:2018aa,Ogata:2020ab,Ogata:2020aa,Yao:2020PRX,Yao:2021aa,Su:2024aa}. 

%However,
%identifications of gaplessness and gappedness are 
However, theoretical determination of the presence or absence of an excitation gap
in many-body systems can be a notoriously difficult task because the ``gapped'' and ``gapless'' phases are concepts defined in the thermodynamic limit;
there is no algorithm to determine the spectral gap for a general many-body spin system~\cite{bausch2020undecidability}.  
% Unfortunately, it typically demands several times of system size to finish especially under periodic boundary conditions (PBCs). 
A useful clue is provided by the famous Lieb-Schultz-Mattis (LSM) theorem~\cite{Lieb:1961aa} and its various extensions~\cite{OYA1997,Oshikawa:2000aa,Hastings:2004ab,NachtergaeleSims}, 
stating that one-dimensional quantum spin models with a half-integral spin per unit cell cannot be gapped with a \textit{unique} ground state in the presence of SO(3) spin-rotation symmetry and lattice translation symmetry.
Nevertheless,
the LSM-type theorems have two shortcomings.
(a)~Since they are no-go type theorems,
they cannot make definitive statements when a required microscopic condition is not satisfied by lattice spin models (e.g., spin-1 or integral spin chains).
(b)~Even for a lattice model satisfying the microscopic conditions required by the LSM-type theorems,
they cannot distinguish gapless models from gapped models with \textit{degenerate} ground states.
To resolve these conceptual and substantial difficulties remains an open question.

%These limitations motivate the search for alternative diagnostics both robust and computationally efficient. 
%One promising direction involves nonlocal observables derived from topological considerations. 
%Notably, Resta proposed a formulation of electric polarization to characterize conduction properties
%in periodic one-dimensional systems~\cite{Resta:1998aa,Resta:1999aa,Aligia:1999aa}, 
%which corresponds to the Lieb-Schultz-Mattis (LSM) twisting operator in spin chains~\cite{Lieb:1961aa}:
%\begin{eqnarray}
%\hat{U}= \exp \left( \frac{2 \pi i}{L} \sum_{j = 1}^{L} j \hat{S}^z_j \right),   
%\end{eqnarray}
%originally introduced in the proof of the LSM theorem~\cite{Lieb:1961aa,Affleck:1986aa,Oshikawa:2000aa,Hastings:2004ab,NachtergaeleSims} 
%and classification of ingappibilities~\cite{PhysRevB.96.195105,PhysRevLett.118.021601,Yao:2019aa}. 
%% The twisting operator encodes momentum shifts under twisted boundary conditions and plays a central role in recent theoretical developments.
%Prior studies have shown that it captures both the sign~\cite{Tasaki:2018aa} and quantized value~\cite{Su:2024aa} of topological indices, 
%and also successfully detects various valence-bond-solid (VBS) states numerically~\cite{Nakamura:2002aa}. 
%However, its applicability to systems with degenerate ground states remains largely unexplored, 
%posing an open question in the study of intrinsic energy spectrum. 
% This represents a critical gap in our understanding of the relationship between nonlocal observables and spectral properties.

In this Letter, 
we present a general, necessary condition for spin chains with spin-rotation symmetry
to be gapped, and show that it can be used to identify gapless phases.
%propose a rigorous way to sufficiently identify gapless phases for general spin-rotation symmetric spin chains by stating necessary conditions for it to be gapped.
Important ingredients for our discussion are the twisting operator~\cite{Lieb:1961aa,OYA1997,Resta:1998aa,Resta:1999aa,Aligia:1999aa,Oshikawa:2000aa,Nakamura:2002aa,Hetenyi:2019aa,Hetenyi:2020aa,Aligia:2023aa,Su:2024aa,Tasaki:2018aa}
\begin{eqnarray}
\hat{U}= \exp \left( \frac{2 \pi i}{L} \sum_{j = 1}^{L} j \hat{S}^z_j \right)
\end{eqnarray}
and the ground-state expectation value of its square, $\langle\hat{U}^2\rangle$.
We rigorously prove that 
% $\langle\hat{U}^2\rangle=1+\mathscr{O}(1/\sqrt{L})\rightarrow1$ 
$\langle\hat{U}^2\rangle=1+\mathscr{O}(1/L)\rightarrow1$
for any ground state of gapped SO(3)-symmetric spin chains with extended translation symmetry (\textit{Theorem~5}).
We note that, unlike the LSM theorem,
our theorem does not require stringent conditions, besides the SO(3) symmetry, on microscopic data per unit cell of spin chains,
so that it is applicable to spins of any size $S$.
%More essentially,
Furthermore,
it is numerically tractable since an \textit{arbitrarily} chosen ground state can be used to evaluate $\langle \hat{U}^2\rangle$.
%and its convergence rate is completely under control and 
%where the scaling of finite-size corrections are estimated.
We perform density matrix renormalization group (DMRG)~\cite{White:1992aa,White:1993aa,itensor} 
and unbiased Quantum Monte Carlo (QMC)~\cite{Sandvik:1991aa,Sandvik:1992aa,Syljuasen:2002aa,kawashima:2004aa,Gubernatis:2020,sadoune:2022,Zhang:2024aa} simulations
for several spin models,
and present both evidence for (potentially) gapped phases and identification of gapless phases, which is ensured by reliable finite-size scaling analysis.

% We perform density matrix renormalization group (DMRG) simulations for several spin models~\cite{White:1992aa,White:1993aa,itensor}, 
% and present both evidence for (potentially) gapped phases and identification of gapless phases,
% which is ensured by reliable finite-size scaling analysis (with truncation errors below $10^{-7}$).

\paragraph{Symmetry, locality, and extended translation.---}
To set the stage, 
we introduce various notations and concepts that are essential in our discussion of spin chains.
A spin-chain Hamiltonian $\mathcal{H}$ is defined on a one-dimensional lattice of length $L$,
where spin $\bm{S}_j=(\hat{S}^x_j,\hat{S}^y_j,\hat{S}^z_j)$ acts on the $(2s_j+1)$-dimensional Hilbert space on site $j$, under periodic boundary conditions (PBCs), $\bm{S}_{j+L}=\bm{S}_j$. 
The Hamiltonian possesses $\text{U}(1)\rtimes \mathbb{Z}_{2}\subset\mathrm{SO}(3)$ spin-rotation symmetry;
the full SO(3) symmetry will be imposed later (from \textit{Theorem 3}).
Here, $\text{U(1)}$ and $\mathbb{Z}_{2}$ spin-rotation symmetries are generated by
$\hat{S}^z_\mathrm{tot}$ and $\R\equiv\exp(i\pi\hat{S}^x_\mathrm{tot})$, respectively,
where $\hat{S}^\alpha_\mathrm{tot}=\sum_{j=1}^L\hat{S}^\alpha_j$ ($\alpha=x,y,z$).
%They satisfy
%\begin{eqnarray}\label{SU2_relationship}
%\hat{S}_j^z+R_\pi^x\hat{S}_j^z{\left(R_\pi^x\right)}^{-1}=0.
%\end{eqnarray}

The Hamiltonian is assumed to be \textit{local}, 
i.e., it admits a decomposition  
$\mathcal{H}=\sum_{j=1}^{L}h_j$,
where each $h_j$ acts as the identity operator on the spins at a distance larger than $d_L$ from site $j$,
with $d_L$ being upper bounded by some model-dependent $d$ for $L\gg1$.  %in the thermodynamic limit.
Moreover, the local decomposition can always be chosen to respect the on-site symmetry~\cite{Su:2024aa},
\begin{eqnarray}\label{local_sym}
[\hat{S}^z,h_j]=[R^x_\pi,h_j]=0.
\end{eqnarray}

%Moreover, to make the thermodynamic limit mathematically well-defined, 
Furthermore, the Hamiltonian is assumed to be invariant under a translation $\hat{T}$  
\begin{eqnarray}\label{OperatorForm_T}
\hat{T}^{-1} \bm{S}_j \hat{T} = \bm{S}_{j+r},
\end{eqnarray}
where an integer $r$ is a model-dependent minimum period,
so that we can increase the system size $L$ as $L=Nr$ with $N\rightarrow\infty$ to define the thermodynamic limit.
We call the translation by the minimum period $r$ \textit{extended translation}. %\textit{extended translation}. %\textit{extensibility}.
There is a useful relationship between $\hat{U}$ and $\hat{T}$: 
\begin{eqnarray}\label{TUT}
\hat{T}^{-1} \hat{U} \hat{T}=\hat{U} 
\exp\!\left( 2 \pi i \sum_{j = 1}^{r} \hat{S}^z_{j} \right) 
\exp\!\left(- \frac{2 \pi i}{L} r \hat{S}^z_\mathrm{tot} \right), 
\end{eqnarray}
where $\exp \!\left( 2 \pi i \sum_{j = 1}^{r} \hat{S}^z_{j} \right)\in\{\pm1\}$ is simply
a $\mathbb{Z}_2$ phase.
% a U(1) phase. 
One should note that the period $r$ can be any integer and model-dependent, and
we do not need to impose translation symmetry with $r=1$ or any other microscopic constraint on each unit cell, unlike the LSM theorem.

\textit{Gapped and gapless Hamiltonians.---}
For the sake of rigor, we introduce the concept of ``gapped'' and ``gapless'' following Refs.~\cite{zeng2019quantum,Zeng:2015aa}.
The system is said to be \textit{gapped} in the limit of the system size $L\rightarrow\infty$,
when the following condition is met:
the ground-state degeneracy $g_L$ of the $L$-series of Hamiltonians $\mathcal{H}_L$ is upper bounded by a finite integer $g$,
and the energy gap $\Delta_L$ between the $g_L$ ground state(s) and the first excited states of $\mathcal{H}_L$ is lower bounded by a finite positive number $\Delta$.
%Then we call the system is \textit{gapped}.
Otherwise,
the system is \textit{gapless}.

As preparations for our main statements,
we prove several Lemmas.

\paragraph{Lemma~1.---}
If a spin-chain Hamiltonian $\mathcal{H}$ has $[\text{U(1)}\rtimes\mathbb{Z}_2]$-symmetry,
then,
for \textit{any} $(\R,\mathcal{H})$-eigenstate $|\psi\rangle$ with energy $E$,
the ``twisted'' state $|\Psi\rangle\equiv\hat{U}|\psi\rangle$ has its energy expectation value within $\mathscr{O}(1/L)$ energy window around $E$.

\textit{Proof:}
Following the authors' previous work~\cite{Su:2024aa},
we calculate $\Delta E\equiv\langle\Psi|\mathcal{H}|\Psi\rangle-\langle\psi|\mathcal{H}|\psi\rangle$
%by defining a more general twisting operator 
and use the locality of $\mathcal{H}$ and the U(1) symmetry of $h_j$ to obtain
%\begin{eqnarray}
%\hat{U}^{(j)}\equiv\exp \left( \frac{2 \pi i}{L} \sum_{m = j+1}^{j+L} m \hat{S}^z_m \right)
%\end{eqnarray} 
%in region $j\in[-L,L)$.
%It is equals to $\hat{U}$ up to a trivial U(1) phase~\cite{Ugeneral}.
% \begin{eqnarray}
% \hat{U}^{(j)}&\equiv&\exp \left( \frac{2 \pi i}{L} \sum_{m = j+1}^{j+L} m \hat{S}^z_m \right)\nonumber\\
% &=&\left\{\begin{array}{ll}\hat{U}\exp\left(\sum_{n=1}^j2\pi i\hat{S}^z_n\right),&\text{ if }j> 0;\\
% \hat{U},&\text{ if }j= 0;\\
% \hat{U}\exp\left(-\sum_{n=j}^{-1}2\pi i\hat{S}^z_n\right),&\text{ if }j< 0,\end{array}\right.
% \nonumber\\
% &\propto&\hat{U},
% \end{eqnarray}
%The energy difference becomes
%\begin{eqnarray}
%\Delta E=\sum_{j=1}^{L} \left\langle \psi\bigg| \hat{U}^{(j-2d) \dagger} h_j  \hat{U}^{(j-2d)} - h_j \bigg| \psi\right\rangle. 
%\end{eqnarray}
%Then we add a U(1) phase $\exp(-2\pi i\hat{S}_j^z j/L)$ on each site $j$ which does nothing, 
%and the locality of $h_j$ to find
%\begin{eqnarray}
%\Delta E&=& \sum_{j=1}^{L} \left\langle \psi\bigg| \exp \left[- \frac{2 \pi i}{L} \sum_{|m-j|\leq d} (m-j) \hat{S}^z_m \right] h_j \right. \nonumber\\ 
%&&\times \left. \exp \left[ \frac{2 \pi i}{L} \sum_{|n-j|\leq d} (n-j) \hat{S}^z_n \right] - h_j \bigg| \psi\right\rangle. 
%\end{eqnarray}
%It is observed that both the terms of $h_j$ and $(m-j)$, $(n-j)$ are bounded even under limit $L\rightarrow\infty$, 
%thereby one can expand the exponentials as Taylor series by $(1/L)$'s
\begin{eqnarray}
\Delta E\!=\! \sum_{j=1}^{L} \left\langle \psi\bigg|
 \left[h_j, \frac{2 \pi i}{L} \!\!\sum_{|m-j|\leq d} (m-j) \hat{S}^z_{m} \right]
  \bigg| \psi\right\rangle\! +\!\mathscr{O}\!\left(\frac{1}{L}\right)\!.\nonumber
%&&+ \sum_{j=1}^{L} \langle \psi| \mathscr{O} (1 / L^2) | \psi\rangle.
\end{eqnarray}
Since $h_j$ is $\R$-symmetric and $|\psi\rangle$ is a $\R$-eigenstate, we can 
replace $\hat{S}^z_{m}$ by $\left[\hat{S}_m^z+R_\pi^x\hat{S}_m^z{\left(R_\pi^x\right)}^{-1}\right]/2$, which vanishes.
This completes the proof.

% , so the proof of \textit{Lemma 1} is completed.

\textit{Lemma~2.---}
For any eigenstate $|\psi_\text{gs}\rangle$ in the ground-state sector of a gapped spin-chain Hamiltonian $\mathcal{H}$ with U(1)$\rtimes\mathbb{Z}_2$ symmetry,
the twisted state $\hat{U}|\psi_\text{gs}\rangle$ can be written as
\begin{eqnarray}
\hat{U}|\psi_\text{gs}\rangle=|\Psi_\text{gs}\rangle+|\xi\rangle, 
\end{eqnarray}
where $|\Psi_\text{gs}\rangle$ is a state in the ground-state sector with almost a unit norm  
and $|\xi\rangle$ is an excited state with nearly a zero norm, i.e.,
\begin{eqnarray}
\langle\Psi_\text{gs}|\Psi_\text{gs}\rangle=1+\mathscr{O}\!\left(\frac{1}{L}\right)\!,
\,\,\,
% &&\langle\xi|\Psi_\text{gs}\rangle=0,\\
\langle\xi|\xi\rangle=\mathscr{O}\!\left(\frac{1}{L}\right)\!,
\,\,\,
\langle\xi|\Psi_\text{gs}\rangle=0.
\nonumber
\end{eqnarray}

% \paragraph{Proof---}
\textit{Proof: }We take an orthonormal basis $\{|\text{gs}_j\rangle | {j=1,2,\cdots,g_L}\}$ of the ground-state sector, where $|\text{gs}_j\rangle$ are
simultaneously $\R$-eigenstates.
We write $|\psi_\text{gs}\rangle$ as
\begin{eqnarray}
|\psi_\text{gs}\rangle=\sum_{j=1}^g\gamma_j|\text{gs}_j\rangle,
\quad
\sum_{j=1}^g|\gamma_j|^2=1.
\end{eqnarray}
For each $|\text{gs}\rangle_j$, \textit{Lemma~1} implies 
$\hat{U}|\text{gs}_j\rangle=|\Psi_\text{gs$j$}\rangle+|\xi_j\rangle$.  
Here it is important to note that the existence of a nonzero gap $\Delta$ ensures that
the ground-state component $|\Psi_\text{gs$j$}\rangle$ has a norm
$\langle\Psi_\text{gs$j$}|\Psi_\text{gs$j$}\rangle = 1 + \mathscr{O}(1/L)$
while $|\xi_j\rangle$ above the ground state(s) has a vanishing norm
$\langle\xi_j|\xi_j\rangle=\mathscr{O}(1/L)$~\cite{tada:2025aa}. 
Thus
\begin{eqnarray}
\hat{U}|\psi_\text{gs}\rangle=\sum_{j=1}^g\gamma_j\left(|\Psi_{\text{gs}j}\rangle+|\xi\rangle_j\right)\equiv|\Psi_\text{gs}\rangle+|\xi\rangle,
\end{eqnarray}
where the first term on the right-hand side gives a state in the ground-state sector, 
and the norm of second term is given by 
$\langle\xi|\xi\rangle=\sum_{j,k=1}^g\gamma_j^*\gamma_k\langle\xi_j|\xi_k\rangle=\mathscr{O}(1/L)$. 
Consequently, the unitarity of $\hat{U}$ gives 
$\langle\Psi_\text{gs}|\Psi_\text{gs}\rangle=1+\mathscr{O}(1/L)$ completing the proof. 

\textit{Theorem~3: Singlet criterion.---}
If spin-chain Hamiltonian $\mathcal{H}$ with extended translation symmetry
and SO(3) spin-rotation symmetry is gapped, then
any eigenstate in its ground-state sector
is a spin singlet state $\bm{\hat{S}}_\mathrm{tot}=0$. 

\textit{Proof:}
This theorem can be proved by contradiction.  
Let us assume a non-singlet spin representation
in the ground-state sector, which must include a ground state
$|\text{gs}\rangle$ staisfying $\hat{S}^z_\mathrm{tot}|\text{gs}\rangle=m|\text{gs}\rangle$
with $m = 1/2$ or $1$. 
Since both $\mathcal{H}$ and $\hat{S}^z_\mathrm{tot}$ commutes with the translation operator $\hat{T}$, 
the state $|\text{gs}\rangle$ can be chosen to satisfy  
\begin{eqnarray}
&&\hat{T}|\text{gs}\rangle=\exp(iP_0)|\text{gs}\rangle,
\end{eqnarray}
where $P_0$ is its lattice momentum. 
Equation~\eqref{TUT} gives 
\begin{eqnarray}
\hat{T}\hat{U}|\text{gs}\rangle=\exp[i(P_0+\Delta P)]\hat{U}|\text{gs}\rangle,
\end{eqnarray}
where $\Delta P$ takes a definite value in the following set: 
\begin{eqnarray}\label{DeltaP}
\Delta P \in \left\{- \frac{\pi r}{L}, - \frac{2\pi r}{L}, \pi - \frac{\pi r}{L}, \pi - \frac{2\pi r}{L}\right\}.
\end{eqnarray}
Thus, the state $\hat{U}|\text{gs}\rangle$ has the lattice momentum $P_0+\Delta P$
and is a normalized state in the ground-state sector in the thermodynamic limit,
according to \textit{Lemma 2}. 
Repeating this procedure generates the states
$\hat{U}^q|\text{gs}\rangle$ for $q=1, 2, 3, \cdots,g,g+1,\cdots$, 
a sequence of states within $\mathscr{O}(1/L)$-energy window around $|\text{gs}\rangle$
with distinct momenta $P_0+q\Delta P$,
regardless of the specific value $\Delta P$ taken in Eq.~\eqref{DeltaP},
when,
e.g., $L\gg2gr$. 
As a result, the number of these states in the ground-state sector must exceed $g$ in the thermodynamic limit. 
It contradicts the assumption of upper-bounded ground-state degeneracy
in the definition of gapped Hamiltonians. 
This completes the proof of \textit{Theorem~3}. 

\textit{Theorem~3}
provides a powerful criterion for determining gaplessness in SO(3)-symmetric spin chains with extended translation symmetry. 
For instance, it directly ensures the gaplessness of half-integer spin chains of odd lengths, since such systems do not admit singlet states in its Hilbert space.

\paragraph{Corollary~4.---}
Any SO(3)-symmetric spin chain with half-integral spin on each site with an extended periodicity must be gapless if the thermodynamic limit is taken as $L=2N+1$ with $N\to\infty$.

The consideration that the entire odd-$L$ half-integer chain,
as a zero-dimensional quantum mechanical system,
forms a projective representation of SO(3) symmetry only implies that the ground-state energy (actually any energy) must be at least doubly degenerate.
The gaplessness imposed by \textit{Corollary~4} is genuinely a many-body effect beyond such a zero-dimensional single-body viewpoint.
%{\bf (AF: What happens if spins are not interacting?)} 
%{\color{red}\bf YY: If all spins don't interact with each other, the GS-deg is $2^L$, which is gapless by our definitions of ``being gapped'', and ``not being gapped''=``gapless''.}

%So far,
%we have developed adequate ingredients to propose and prove our main theorem.
With these preparatory Corollaries and Theorems, we are now ready to propose and prove our main theorem.

\paragraph{Theorem 5: $\langle\hat{U}^2\rangle$-criterion.---} 
If a spin chain with SO(3) spin-rotation symmetry and extended translation symmetry is gapped,
then any state in its ground-state sector,
say $|\text{GS}\rangle$,
%is an SU(2) singlet and
satisfies that
\begin{eqnarray}\label{Theorem5}
\langle\text{GS}|\hat{U}^2|\text{GS}\rangle=1+
\mathscr{O}(1/L)\rightarrow1.
\end{eqnarray}

\textit{Proof:} 
\textit{Lemma~2} implies that
\begin{eqnarray}\label{Theorem5-0}
\hat{U}|\text{GS}\rangle=|\Psi_\text{GS}\rangle+|\gamma\rangle,
\end{eqnarray}
with $\langle\Psi_\text{GS}|\Psi_\text{GS}\rangle=1+\mathscr{O}(1/L)$ 
and $\langle \gamma|\gamma\rangle=\mathscr{O}(1/L)$. 
It follows from the identity $\hat{U}\R \hat{U}(\R)^{-1}=1$ that
\begin{eqnarray}\label{t5_0}
1=\langle\text{GS}|\hat{U}\R \hat{U}(\R)^{-1}|\text{GS}\rangle.
\end{eqnarray}
Since any state in the ground-state sector is an SU(2) singlet by \textit{Theorem 3},
$\R|\text{GS}\rangle=(\R)^{-1}|\text{GS}\rangle=|\text{GS}\rangle$ and
$\R|\Psi_\text{GS}\rangle=(\R)^{-1}|\Psi_\text{GS}\rangle=|\Psi_\text{GS}\rangle$.
Thus,
Eq.~(\ref{t5_0}) becomes
\begin{eqnarray}\label{Theorem5-1}
1&=&\langle\text{GS}|\hat{U}\R (|\Psi_\text{GS}\rangle+|\gamma\rangle)\nonumber\\
&=&\langle\text{GS}|\hat{U}|\Psi_\text{GS}\rangle+\langle\text{GS}|\hat{U}R_{\pi}^x|\gamma\rangle. 
\end{eqnarray}
On the other hand, we have
\begin{eqnarray}
\langle\text{GS}|\hat{U}^2|\text{GS}\rangle=\langle\text{GS}|\hat{U}|\Psi_\text{GS}\rangle+\langle\text{GS}|\hat{U}|\gamma\rangle,
\end{eqnarray}
and,
by using Eq.~\eqref{Theorem5-1}, we find
\begin{eqnarray}
\langle\text{GS}|\hat{U}^2|\text{GS}\rangle&=&1-\langle\text{GS}|\hat{U}R_{\pi}^x|\gamma\rangle+\langle\text{GS}|\hat{U}|\gamma\rangle \nonumber\\
&=&1-\langle\text{GS}|\hat{U}^{\dagger}|\gamma\rangle+\langle\text{GS}|\hat{U}|\gamma\rangle. 
\end{eqnarray}
By replacing $\hat{U}$ by $\hat{U}^{-1}=\hat{U}^{\dagger}$ in \textit{Lemma 2}, one has $\hat{U}^{\dagger}|\text{GS}\rangle=|\Phi_\text{GS}\rangle+|\zeta\rangle$
% similar with Eq.~\eqref{Theorem5-0}, then 
% we obtain
, then 
\begin{eqnarray}
\langle\text{GS}|\hat{U}^2|\text{GS}\rangle=1-\langle\gamma|\gamma\rangle + \langle\zeta|\gamma\rangle=1+\mathscr{O}(1/L), 
% \rightarrow 1, 
\end{eqnarray}
where we use 
$\langle\zeta|\gamma\rangle\leq\sqrt{\langle\zeta|\zeta\rangle\langle\gamma|\gamma\rangle}$.  
It completes the proof of \textit{Theorem~5}.

\textit{Theorem~5} can be formally generalized by defining
\begin{eqnarray}
\hat{U}_{\bm{n}}\equiv\exp\left[i\frac{2\pi}{L}\sum_{j=1}^Lj\left(\bm{n}\cdot\bm{S}_j\right)\right],
\end{eqnarray} 
with a unit 3-vector $\bm{n}$~\footnote{One should note that $\hat{U}_{\bm{n}+\bm{m}}\neq \hat{U}_{\bm{n}}\hat{U}_{\bm{m}}$ in general.}.
For any $\bm{n}$,
there exists an SU(2) spin-rotation transformation $R_{\bm{n}}$ such that
$\hat{S}^z_j=R_{\bm{n}}\left(\bm{n}\cdot\bm{S}_j\right)R_{\bm{n}}^{-1}$ or $\hat{U}=R_{\bm{n}}\hat{U}_{\bm{n}}R_{\bm{n}}^{-1}$.
Noticing $R_{\bm{n}}|\text{GS}\rangle=|\text{GS}\rangle$ by \textit{Theorem 3} and the unitarity of $\hat{U}_{\bm{n}}$ and $R_{\bm{n}}$,
we obtain the following formal extension.

\textit{Corollary~6.---}Under the same condition of \textit{Theorem~5},
\begin{eqnarray}
\langle\text{GS}|\hat{U}_{\bm{n}_1}\hat{U}_{\bm{n}_2}\cdots\hat{U}_{\bm{n}_{p}}|\text{GS}\rangle=1+\mathscr{O}(1/L),
\end{eqnarray}
for any even $p\in2\mathbb{Z}$ and unit 3-vectors $\bm{n}_{1,2,\cdots,p}$.

\paragraph{Identifying gaplessness and evidence of gappedness.---}Since \textit{Theorem~5} provides a necessary quantitative condition for an SO(3)-symmetric Hamiltonian to be gapped,
it can be utilized to identify gapless Hamiltonian as well as provide supporting evidence for various potentially gapped Hamiltonian. 

% \begin{figure}[tbp]
% \centering
% \subfigure{\includegraphics[width=0.2385\textwidth]{data_MGeven.pdf}}
% \hfill
% \subfigure{\includegraphics[width=0.2385\textwidth]{data_MGodd.pdf}}
%  \caption{\justifying
% Results for $\langle\hat{U}^2\rangle$ in the Majumdar-Ghosh model. 
% (a) For even chain lengths, DMRG data correspond to the first ground state at each system size $L$. 
% The expectation values scale linearly versus $1/L$ [see Eq.~\eqref{Usquare_MGmodel}], with intercept of 0.996(1) in the thermodynamic limit. 
% The analytical results for two exact ground states are also showed into pink and blue lines (see Supplementary Materials~\cite{Append}). 
% (b) For odd chain lengths, the expectation value remains below $10^{-4}$ for all system sizes considered.
% }
% \label{fig:MG}
% \end{figure}

\begin{figure}[tbp]
\centering
\subfigure{\includegraphics[width=0.2385\textwidth]{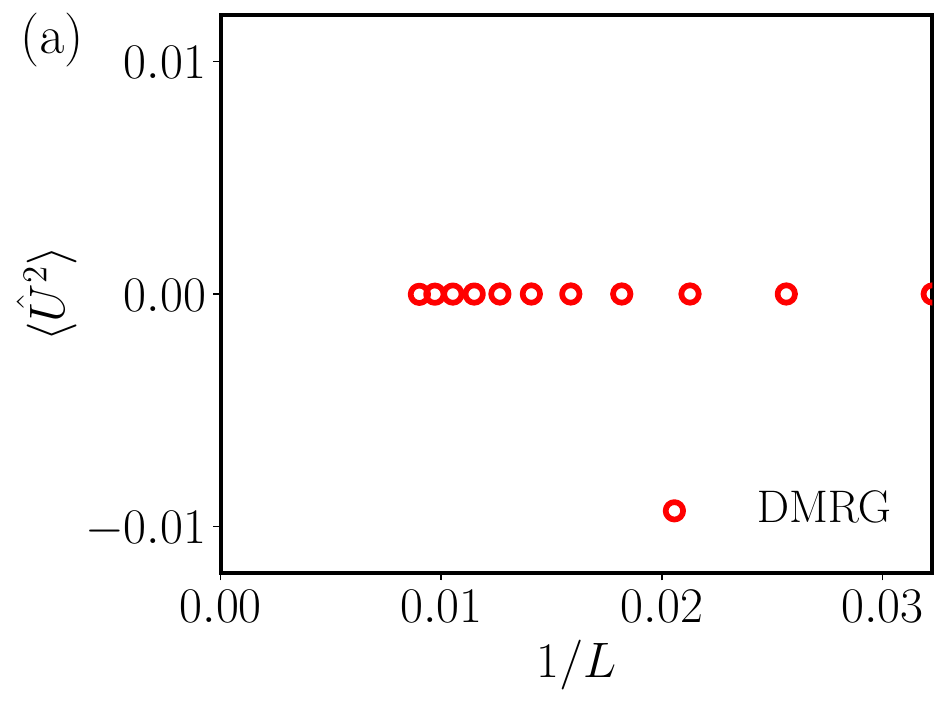}}
\hfill
\subfigure{\includegraphics[width=0.2385\textwidth]{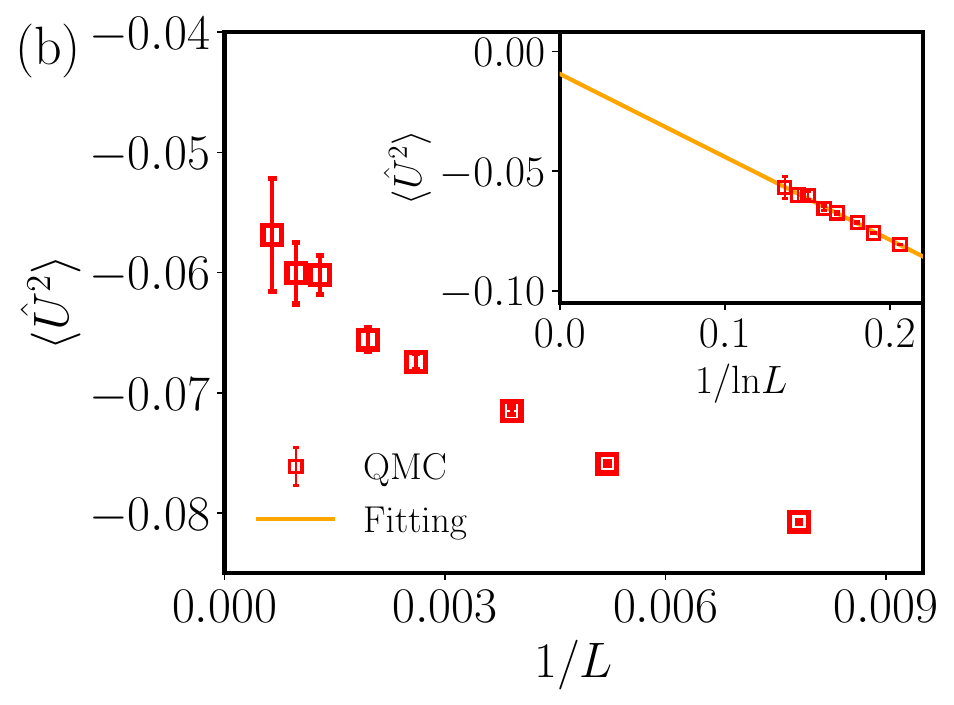}}
\caption{\justifying
Results for $\langle\hat{U}^2\rangle$ in the gapless spin-1/2 models: Majumdar-Ghosh model with odd chain lengths (left) and Heisenberg antiferromagnetic chain (right). 
Both show the tendency of $\langle\hat{U}^2\rangle \ll 1$. 
The latter dataset is fitted by $\langle\hat{U}^2\rangle = a - b/\ln L$, with $a=-0.009(4)$, $b=0.346(7)$, shown in the inset.
}
\label{fig:spin-1/2-gapless}
\end{figure}

Let us begin with the $S=1/2$ Majumdar-Ghosh (M-G) model
$\mathcal{H}_{\text{M-G}}=\sum_{j=1}^{L} \bm{S}_j \cdot \bm{S}_{j+1}+\frac{1}{2}\sum_{j=1}^{L}\bm{S}_j \cdot \bm{S}_{j+2}$ with length $L$ and PBCs~\cite{10.1063/1.1664978,10.1063/1.1664979,Majumdar1970,Shastry:1981aa,Caspers1984,Affleck:1988ab,Chhajlany:2007aa}. 
For even chain lengths $L\in2\mathbb{Z}$, its lowest-energy sector is doubly degenerate and either of them satisfies~\cite{Append}
\begin{eqnarray}\label{Usquare_MGmodel} 
\langle\hat{U}^2\rangle
= 1-\frac{\pi^2}{L}+\mathscr{O} \left(\frac{1}{L^2}\right).
\end{eqnarray}
By contrast, for odd chain lengths $L\in2\mathbb{Z}+1$~\cite{Affleck:1988ab}, there is no available exact ground-state wavefunction. 
The DMRG simulations (under the PBCs, with trunctaion error no more than $10^{-7}$-order) reveal that 
$\langle\hat{U}^2\rangle$
remains below $10^{-4}$ for all sizes under consideration as in FIG.~\ref{fig:spin-1/2-gapless}~(a),  
indicating the gaplessness. 
It is in accordance with the conclusion from \textit{Corollary~4} that odd-$L$ MG chains are gapless. 

{
We also apply the $\langle\hat{U}^2\rangle$-criterion to spin-$S$ Heisenberg antiferromagnetic (HAF) chains 
$\mathcal{H}_{\text{HAF}}=\sum_{j=1}^L\bm{S}_j \cdot \bm{S}_{j+1}$}. 
We calculate $\langle\hat{U}^2\rangle$ for spin-$1/2$ HAF by QMC simulations up to $L=1536$ (the largest inverse temperature $\beta = 6L$) with PBCs;
see FIG.~\ref{fig:spin-1/2-gapless}~(b).
The values of $\langle\hat{U}^2\rangle$ remain close to zero, indicating gaplessness~\cite{Bethe1931,10.1143/PTP.46.401}. 
Moreover, they are well fitted by $\langle\hat{U}^2\rangle = a - b/\ln L$
with $a=-0.009(4)$, $b=0.346(7)$, as shown in the inset of Fig.~\ref{fig:spin-1/2-gapless}~(b).

\begin{figure}[tbp] 
\centering
\subfigure{\includegraphics[width=0.2385\textwidth]{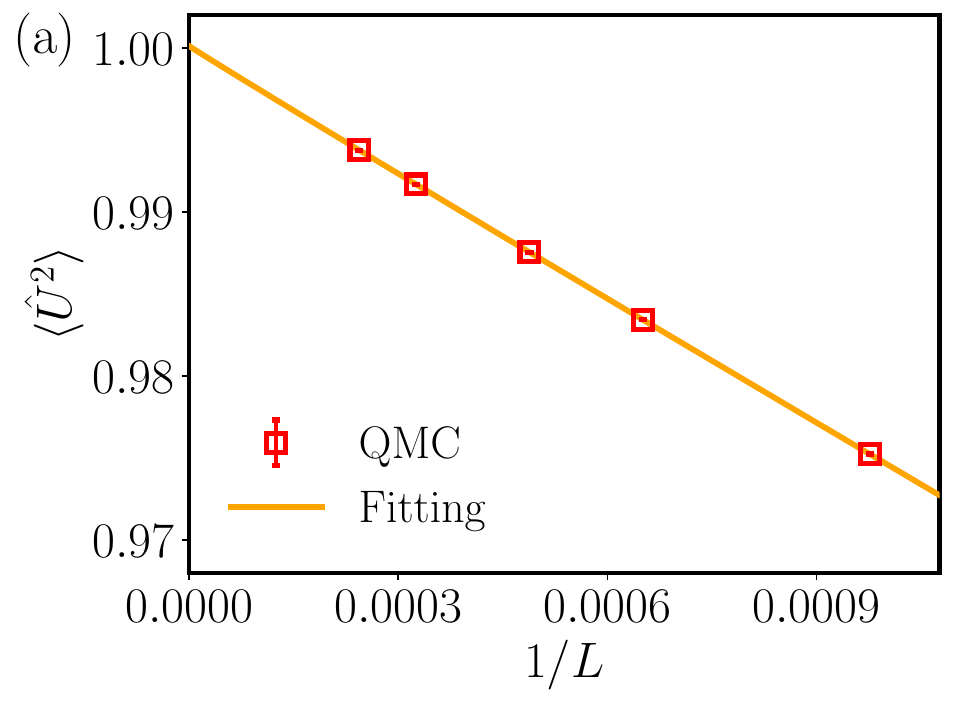}}
\hfill
\subfigure{\includegraphics[width=0.2385\textwidth]{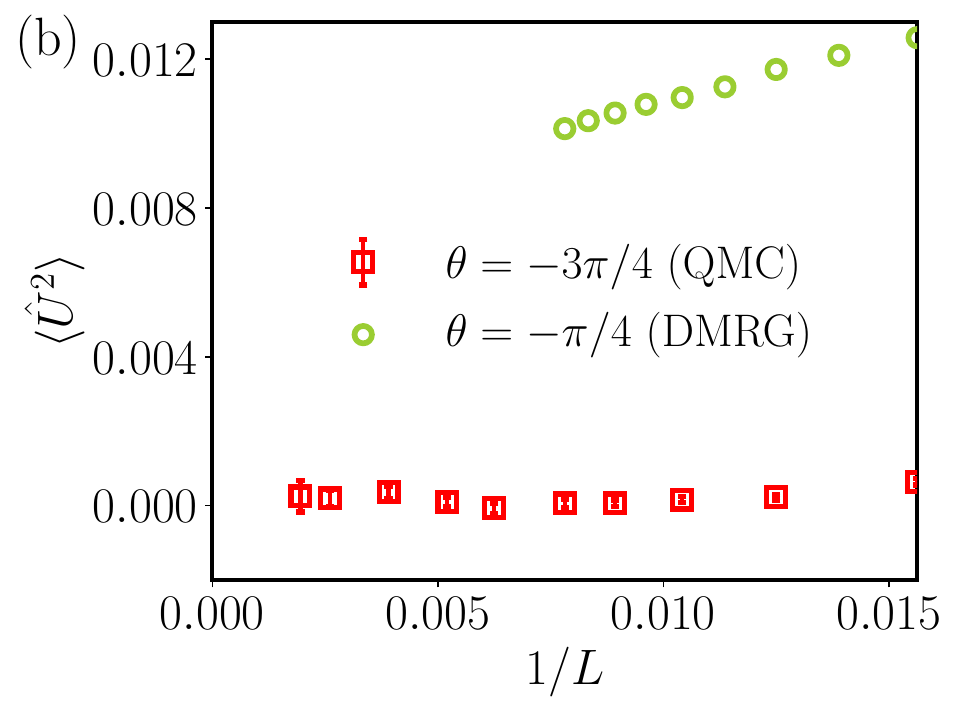}}
\caption{\justifying
Results for $\langle\hat{U}^2\rangle$ as a function of $1/L$ for the BLBQ chains. 
The left is $\theta=0$ case, i.e., spin-1 HAF chain, which is fitted by Eq.~\eqref{DMRG_fitting}, yielding $a=1.000(1)$ and $b = 22.927(7)$, $k = 0.984(9)$. 
The right one is two gapless points, including QMC results for $\theta=-3\pi/4$ case and DMRG results for $\theta=-\pi/4$ case are shown, resepectviely. 
}
\label{fig:BLBQ}
\end{figure}

For spin-$1$ HAF chain,
QMC simulations up to size $L=4096$ with the largest $\beta=L/4$ are fitted by
\begin{eqnarray}\label{DMRG_fitting}
\langle\hat{U}^2\rangle = a - \frac{b}{L^k}, 
\end{eqnarray}
yielding $a=1.000(1)$, with $b = 22.927(7)$, $k = 0.984(9)$ [Fig.~\ref{fig:BLBQ}~(a)], in agreement with the Haldane conjecture~\cite{HALDANE1983,Haldane:1983aa} and other numerical evidences~\cite{Affleck:1986aa,White:1993ab,Nightingale:1986aa,Golinelli:1994aa}.

The spin-$1$ HAF chain can be regarded as $\theta = 0$ point of the spin-1 bilinear-biquadratic (BLBQ) Heisenberg chain~\cite{Lauchli:2006aa},
\begin{eqnarray}\label{BLBQ}
\mathcal{H}={\cos \theta}\sum_{j=1}^{L} \bm{S}_j \cdot \bm{S}_{j+1}+{\sin \theta}\sum_{j=1}^{L}\left(\bm{S}_j \cdot \bm{S}_{j+1}\right)^2. 
\end{eqnarray}
The Affleck-Kennedy-Lieb-Tasaki (AKLT) model~\cite{Affleck:1987ab} corresponds to $\theta = \arctan (1/3)$. 
By matrix product state representation of the AKLT ground state~\cite{Affleck:1987ab,Affleck:1988ab,Fannes:1992aa}, we obtain~\cite{Append}
\begin{eqnarray}\label{AKLT}
\langle\hat{U}^2\rangle&=& \frac{2^L+\sum_{\eta=\pm}\left(\cos\frac{2\pi}{L}+\eta\sqrt{4-\sin^2\frac{2\pi}{L}}\right)^L}{3^L+3(-1)^L}\nonumber\\
&=&1-\frac{\pi^2}{L}+\mathscr{O}\left(\frac{1}{L^2}\right).
\end{eqnarray}
The extrapolation to the $L\to\infty$ limit with the decaying behavior of $\mathscr{O}(1/L)$ confirms the spectral gap~\cite{Affleck:1987ab,Affleck:1988ab}. 

Furthermore, we investigate the following gapless critical points in the BLBQ model under PBCs:   
one SU(3)-symmetric point $\theta = -3\pi/4$~\cite{Lauchli:2006ab,Voll:2015aa}
and the Takhtajan-Babujan point $\theta = -\pi/4$~\cite{Takhtajan:1982aa,Babujian:1982aa,babujian:1983aa,Rakov:2022aa}.
For $\theta = -3\pi/4$, we do the sign-problem-free QMC simulation for maximum $L=512$ (the largest $\beta=4L$), 
while the DMRG simulations are done for $\theta = -\pi/4$ due to the sign problem of QMC~\cite{harada:2001aa,harada:2002aa,Okunishi:2014aa,Voll:2015aa}. 
Both are shown to have $\langle\hat{U}^2\rangle \ll 1$ in Fig.~\ref{fig:BLBQ} (b), 
identifying the gapless spectra.
On the other hand,
at the SU(3) point $\theta=\pi/4$,
so-called the Uimin-Lai-Sutherland point~\cite{uimin1970one,Lai:1974aa,kulish1981generalized,Sutherland:1975aa,Mashiko:2023aa}, 
$\langle\hat{U}^2\rangle$ from DMRG simulations ranges less than 0.04 indicating gaplessness
with the truncation error only less than $2\times10^{-5}$;
such numerical results for $\theta=\pi/4$ need to be confirmed in the future work.

\paragraph{Conclusions and discussions.---}
In this work, we present a rigorous criterion for gaplessness of spin-rotation symmetric spin chains
by proving that, for any gapped Hamiltonian, (a) its ground state(s) must be SU(2)-singlet; 
(b) the ground-state expectation value of $\langle\hat{U}^2\rangle$ converges to unity in the thermodynamic limit with the finite-size corrections upper bounded by $\mathscr{O}(1/L)$.

\begin{figure}[tbp] 
\centering
\subfigure{\includegraphics[width=0.2385\textwidth]{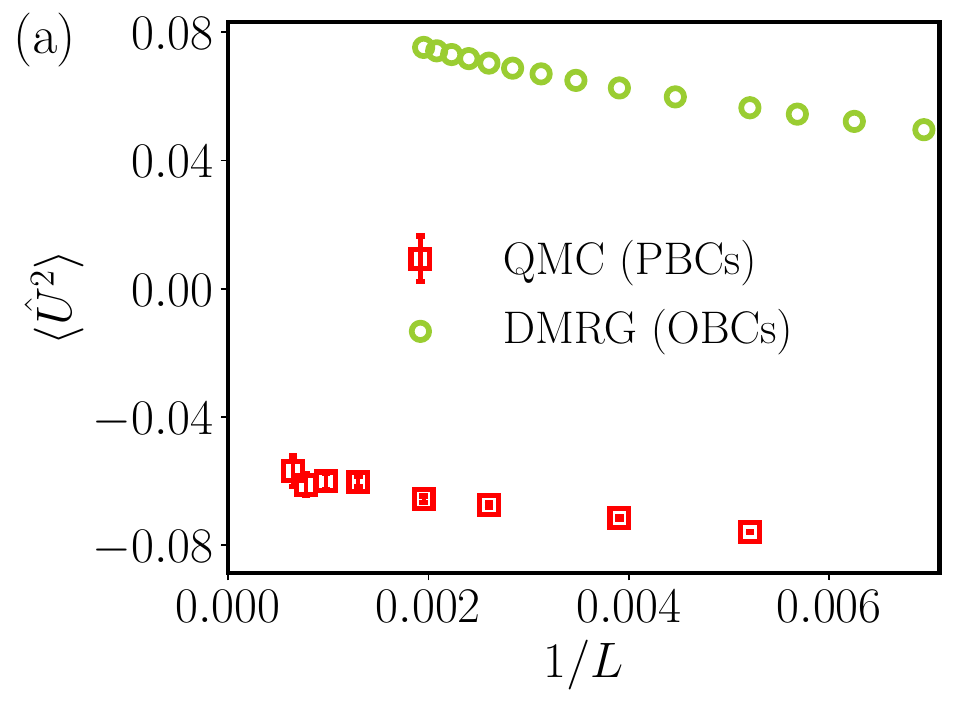}}
\hfill
\subfigure{\includegraphics[width=0.2385\textwidth]{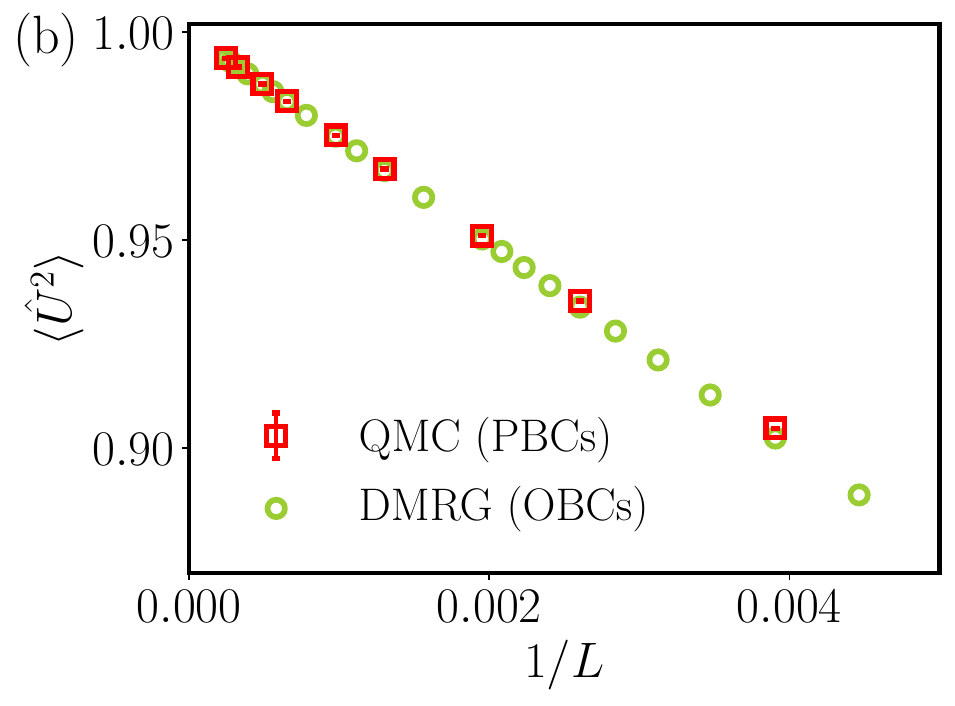}}
\caption{\justifying
Results for $\langle\hat{U}^2\rangle$ as a function of $1/L$ for spin-$1/2$ (left) and spin-$1$ (right) HAF chains, with both OBCs and PBCs.
}

\label{fig:OBCs}
\end{figure}

We apply our theorem to provide supporting evidence for potentially gapped spin models and identify various gapless spin chains by numerics. 
So far,
our theorems and the numerical results are based on PBCs.
Now we apply the DMRG simulations to spin-$1/2$ and spin-$1$ HAF chains under open boundary conditions~(OBCs). 
As shown in FIG.~\ref{fig:OBCs},
$\langle\hat{U}^2\rangle$ of spin-$1/2$ HAF under OBCs is also far less than $1$,
and again the spin-$1$ HAF chain obeys $\langle\hat{U}^2\rangle=1+\mathscr{O}(1/L)$.
The analytical calculations~\cite{Append} of the Majumdar-Ghosh and AKLT models under OBCs lead to the same asymptotic rule $\langle\hat{U}^2\rangle=1+\mathscr{O}(1/L)$ as PBCs.
In the $R^x_\pi$-eigenspace imposing $\langle\hat{U}^2\rangle\in\mathbb{R}$,
even the coefficients of $1/L$ deviations are also the same as PBC results in Eqs.~(\ref{Usquare_MGmodel},\ref{AKLT}),
and a similar behavior is observed numerically in FIG.~\ref{fig:OBCs}~(b).
Therefore,
we expect that
our \textit{Theorem 5} may be true
for a large class of gapped SO(3)-symmetric spin chains under OBCs,
while a mathematically rigorous argument is lacking.
The analytic investigation of our conclusions along this direction,
to be left for future work,
is practically important since OBCs are often better adapted by DMRG than PBCs.

In addition,
our theorem assumes the full SO(3) spin-rotation
symmetry~\cite{Nakamura:2019aa},
which cannot be applied to,
e.g., XXZ models~\cite{Kobayashi:2018ab,Furuya:2019aa}. 
Through the Jordan-Wigner transformation,
the twisting operator $\hat{U}^2$ is mapped to the exponentiated polarization,
and therefore our results might be also understood in spinless fermionic chains~\cite{Hetenyi:2019aa,Hetenyi:2020aa,Aligia:2023aa}.
However,
the boundary effects~\cite{Yao:2021ac} brought by the nonlocal nature of Jordan-Wigner transformations make the correspondence unclear.
Therefore,
the extensions of the current results to spin models with \textit{discrete} spin-rotation symmetry and fermionic systems can be of future interest.

\paragraph{Acknowledgements.---}
The authors are grateful to Yasuhiro Tada for helpful comments on $\mathscr{O}(1/L)$ part in $\langle\hat{U}^2\rangle$ of \textit{Theorem 5} and sharing his manuscript~\cite{tada:2025aa}.
We also thank Linhao Li and Masaki Oshikawa for helpful discussions. 
The work of Y.~Y. was supported by the National Key Research and Development Program of China (Grant No.~2024YFA1408303),
the National Natural Science Foundation of China (Grants No.~12474157, and No.~12447103),
the sponsorship from Yangyang Development Fund,
and Xiaomi Young Scholars Program.
The work of A.F.\ was supported in part by JST CREST (Grant No.~JPMJCR19T2).
The DMRG simulations of this work are performed using the ITensor Julia library (version 0.7)~\cite{itensor}. 
The computations in this Letter were run on the Siyuan-1 and $\pi$2.0 clusters supported by the Center for High Performance Computing at Shanghai Jiao Tong University. 
% \end{acknowledgements}

\paragraph{Data availablity---}
The data that support the findings of this Letter are openly available~\cite{DataAvailbility}.
% https://zenodo.org/records/17052801

\bibliographystyle{apsrev4-1}
%\bibliography{bib}


\newpage
\onecolumngrid 
\appendix
\section{\Large{} Supplemental Materials}

\section{Analytical results of \texorpdfstring{$\langle\hat{U}^2\rangle$}{} for Majumdar-Ghosh model defined on even-length chains}
\subsection{Periodic boundary conditions (PBCs) results}
As stated in the text, there are two degenerated valence-bond-state (VBS) type ground states when the length is chosen as an even number $L\in2\mathbb{Z}$ under PBCs, 
\begin{eqnarray}
|\Psi_{\text{M-G}}^{(1)}\rangle&=&|(1,2)(3,4)\ldots(L-1, L)\rangle,\\
|\Psi_{\text{M-G}}^{(2)}\rangle&=&|(2,3)(4,5)\ldots(L, 1)\rangle,
\end{eqnarray}
where $(m, n)$ denotes singlet localized on sites $m$ and $n$, 
\begin{eqnarray}
|(m,n)\rangle = \frac{1}{\sqrt{2}} \left(|\uparrow_m \downarrow_n\rangle - |\downarrow_m \uparrow_n\rangle\right). 
\end{eqnarray}
One should note that
\begin{eqnarray}
\langle\Psi_{\text{M-G}}^{(2)}|\Psi_{\text{M-G}}^{(1)}\rangle\propto\left(\frac{1}{\sqrt{2}}\right)^{L/2}=\mathscr{O}\left(\frac{1}{L^{+\infty}}\right),
\end{eqnarray}
which enables us to consider that they are orthogonal in the following $L$-scaling.

Since each singlet localized on two different sites is the eigenstate of $\hat{S}^z$ on those sites, 
one can consider the contribution from each singlet $(m, n)$ more generally, for expectation value of $\hat{U}^q$, 
\begin{eqnarray}
\langle(m,n)|\hat{U}^q|(m,n)\rangle&=&\frac{1}{2} \left\{\exp\left[i\frac{2\pi q}{L}\left(\frac{m}{2}-\frac{n}{2}\right)\right]+\exp\left[i\frac{2\pi q}{L}\left(\frac{n}{2}-\frac{m}{2}\right)\right]\right\} \nonumber\\
&=& \cos \!\left(q\pi\frac{m-n}{L}\right), 
\end{eqnarray}
which for our case $q=2$, equals to
\begin{eqnarray}
\langle(m,n)|\hat{U}^2|(m,n)\rangle=\cos\left(2\pi\frac{m-n}{L}\right). 
\end{eqnarray}
Thus we can calculate out for the VBS states 
\begin{eqnarray}
\langle\Psi_{\text{M-G}}^{(1)}|\hat{U}^2|\Psi_{\text{M-G}}^{(1)}\rangle&=&\langle\Psi_{\text{M-G}}^{(2)}|\hat{U}^2|\Psi_{\text{M-G}}^{(2)}\rangle=
\left[\cos\left(\frac{2\pi}{L}\right)\right]^{L/2}=1-\frac{\pi^2}{L}+\mathscr{O}(1/L^2).
\end{eqnarray}
Additionally,
\begin{eqnarray}
&&\langle\Psi_{\text{M-G}}^{(2)}|\hat{U}^2|\Psi_{\text{M-G}}^{(1)}\rangle\propto\left(\frac{1}{\sqrt{2}}\right)^{L/2}=\mathscr{O}\left(\frac{1}{L^{+\infty}}\right);\nonumber\\
&&\langle\Psi_{\text{M-G}}^{(2)}|\Psi_{\text{M-G}}^{(1)}\rangle\propto\left(\frac{1}{\sqrt{2}}\right)^{L/2}=\mathscr{O}\left(\frac{1}{L^{+\infty}}\right).
\end{eqnarray}

Therefore,
for an arbitrary normalized ground state 
$|\Psi_{\text{M-G}}\rangle\equiv\alpha|\Psi_{\text{M-G}}^{(1)}\rangle+\beta|\Psi_{\text{M-G}}^{(2)}\rangle$ spanned by them,
with $|\alpha|^2+|\beta|^2=1+\mathscr{O}(1/L^\infty)$, 
\begin{eqnarray}
\langle\Psi_{\text{M-G}}|\hat{U}^2|\Psi_{\text{M-G}}\rangle
&=&\left(|\alpha|^2\langle\Psi_{\text{M-G}}^{(1)}|\hat{U}^2|\Psi_{\text{M-G}}^{(1)}\rangle+|\beta|^2\langle\Psi_{\text{M-G}}^{(2)}|\hat{U}^2|\Psi_{\text{M-G}}^{(2)}\rangle\right)  \nonumber\\
&=& \left(|\alpha|^2+|\beta|^2\right) \left[1-\frac{\pi^2}{L}+\mathscr{O}\left(\frac{1}{L^2}\right)\right]\nonumber\\
&=&1-\frac{\pi^2}{L}+\mathscr{O}\left(\frac{1}{L^2}\right).
\end{eqnarray}

\subsection{Open Boundary Conditions (OBCs) results}
In the presence of OBCs,
the lowest energy states becomes 5-fold degenerate:
\begin{eqnarray}
|\psi_{\text{M-G}}\rangle&=&|(1,2)(3,4)\ldots(L-1, L)\rangle,\\
|\psi_{\text{M-G}}^{(i,j)}\rangle&=&|(-1)^{i+1},(2,3)(4,5)\ldots(L-2, L-1),(-1)^{j+1}\rangle,\,\,(i,j)\in\{1,2\},
\end{eqnarray}
where there are 4 states with dangling edge spin-$1/2$ excitations.
Calculation of $\langle\hat{U}^2\rangle$ of these five degenerate states satisfies the same rule as PBCs:
\begin{eqnarray}
\langle\psi_\text{M-G}|\hat{U}^2|\psi_\text{M-G}\rangle
&=&1-\frac{\pi^2}{L}+\mathscr{O}\left(\frac{1}{L^2}\right);\nonumber\\
\langle\psi_\text{M-G}^{(i,j)}|\hat{U}^2|\psi_\text{M-G}^{(i,j)}\rangle
&=&\left(\begin{array}{cc}1-\frac{\pi^2}{L}+i\frac{2\pi}{L}&1-\frac{\pi^2}{L}+i\frac{2\pi}{L}\\
1-\frac{\pi^2}{L}-i\frac{2\pi}{L}&1-\frac{\pi^2}{L}-i\frac{2\pi}{L}\end{array}\right)_{i,j}+\mathcal{O}(1/L^2)\nonumber\\
&=&1+\mathcal{O}(1/L).
\end{eqnarray}
Nevertheless,
for a general linear combination,
e.g., two special cases as
\begin{eqnarray}\label{linear_comb}
\left.\begin{array}{c}\left[\frac{1}{\sqrt{2}}\left(\langle\psi_\text{M-G}^{(1,2)}|\pm\langle\psi_\text{M-G}^{(2,1)}|\right)\right]\hat{U}^2\left[\frac{1}{\sqrt{2}}\left(|\psi_\text{M-G}^{(1,2)}\rangle\pm|\psi_\text{M-G}^{(2,1)}\rangle\right)\right]\\\left[\frac{1}{\sqrt{2}}\left(\langle\psi_\text{M-G}^{(1,1)}|\pm\langle\psi_\text{M-G}^{(2,2)}|\right)\right]\hat{U}^2\left[\frac{1}{\sqrt{2}}\left(|\psi_\text{M-G}^{(1,1)}\rangle\pm|\psi_\text{M-G}^{(2,2)}\rangle\right)\right]\end{array}\right\}=1-\frac{\pi^2}{L}+\mathcal{O}(1/L^2),
\end{eqnarray}
which means that the coefficients of $1/L$ are not the same for a general linear combinations,
unlike PBCs cases before.
It is not a coincidence that there is no imaginary part in Eq.~(\ref{linear_comb}) as well as all the PBCs case;
all these states are eigenstates of the global $\pi$-rotation transformation around $x$ axis,
so $\langle\hat{U}^2\rangle$ must be real.

\section{Analytical results of \texorpdfstring{$\langle\hat{U}^2\rangle$}{} for Affleck-Kennedy-Lieb-Tasaki model}
\subsection{PBCs results}
We consider the ground state of AKLT model in translation invariant matrix product states (MPS) form under PBCs~\cite{Affleck:1987ab,Affleck:1988ab,Fannes:1992aa}:
\begin{eqnarray}
\langle\bm{\sigma}|\Psi_{\text{AKLT}}\rangle&=&\sum_{ \alpha_1, \alpha_2, \cdots, \alpha_L\in\{1, 2\}} 
	A_{\alpha_1, \alpha_2}^{\sigma_1} A_{\alpha_2, \alpha_3}^{\sigma_2} \cdots A_{\alpha_L, \alpha_1}^{\sigma_L} \\
	&=& \mathrm{Tr} [\bm{A}^{\sigma_1}\bm{A}^{\sigma_2}\cdots\bm{A}^{\sigma_L}],
\end{eqnarray}
where the vector $\bm{\sigma} = \{\sigma_1, \sigma_2, \cdots, \sigma_L\}$ represents the physical spin-$1$ configuration,
and it can be defined for matrices $\bm{A}^{\sigma}$ in the second line in the $\{0,\pm\}$-basis as
\begin{eqnarray}
\bm{A}^{+}=\left(\begin{array}{cc}
0 & 0 \\
-\frac{1}{\sqrt{2}} & 0
\end{array}\right), \quad \bm{A}^0=\left(\begin{array}{cc}
\frac{1}{2} & 0 \\
0 & -\frac{1}{2}
\end{array}\right), \quad \bm{A}^{-}=\left(\begin{array}{cc}
0 & \frac{1}{\sqrt{2}} \\
0 & 0
\end{array}\right) .
\end{eqnarray}

We note that the VBS state is not normalized, 
which can be calculated out 
\begin{eqnarray}
\langle\Psi_{\text{AKLT}}|\Psi_{\text{AKLT}}\rangle = \mathrm{Tr} [\tilde{\bm{A}}^L] 
= \left(\frac{3}{4}\right)^L + 3\left(-\frac{1}{4}\right)^L, 
\end{eqnarray}
where 
\begin{eqnarray}
\tilde{\bm{A}} = 
\left(\begin{array}{cccc}
\frac{1}{4} & \frac{1}{2} & 0 & 0 \\
\frac{1}{2} & \frac{1}{4} & 0 & 0 \\
0 & 0 & -\frac{1}{4} & 0 \\
0 & 0 & 0 & -\frac{1}{4}
\end{array}\right).
\end{eqnarray}

For the twisting operator, we consider the general case for $\langle\hat{U}^q\rangle$ for any integer $q$, 
\begin{eqnarray}
\hat{U}^q= \exp \left( \frac{2 \pi qi}{L} \sum_{j = 1}^{L} j \hat{S}^z_j \right),
\end{eqnarray}
and define a \textit{site-dependent} parameter for the site-$j$
\begin{equation}
	\theta^{(j)} = 2\pi q j/L. 
\end{equation}
Then, one can yield out for spin-1 system, 
\begin{equation}
	e^{i \theta^{(j)} \hat{S}_j^z} = \mathrm{diag} \{e^{i \theta^{(j)}}, 1, e^{-i \theta^{(j)}}\}. 
\end{equation}

Finally, we deduce out the expectation value for the general twisting operator $\hat{U}^q$
\begin{eqnarray}
\langle\hat{U}^q\rangle= \mathrm{Tr} \left[\prod_{j=1}^{L} \tilde{\bm{C}}^{(j)}\right] \bigg/ \langle\Psi_{\text{AKLT}}|\Psi_{\text{AKLT}}\rangle,
\end{eqnarray}
where 
\begin{eqnarray}
\tilde{\bm{C}}^{(j)} = 
\left(\begin{array}{cccc}
\frac{1}{4} & \frac{1}{2} e^{i \theta^{(j)}} & 0 & 0 \\
\frac{1}{2} e^{-i \theta^{(j)}} & \frac{1}{4} & 0 & 0 \\
0 & 0 & \frac{1}{4} & 0 \\
0 & 0 & 0 & \frac{1}{4}
\end{array}\right).
\end{eqnarray}

We can further simplify it to
\begin{eqnarray}
\langle\hat{U}^q\rangle&=&
\left\{\mathrm{Tr} \left[\prod_{j=1}^{L} \tilde{\bm{D}}^{(j)}\right] 
+ \mathrm{Tr} \left[\prod_{j=1}^{L} \tilde{\bm{E}}^{(j)}\right]\right\} \bigg/ \langle\Psi_{\text{AKLT}}|\Psi_{\text{AKLT}}\rangle \nonumber\\
&=& \left\{\mathrm{Tr} \left[\prod_{j=1}^{L} \tilde{\bm{D}}^{(j)}\right] 
+ \left(\frac{2}{4}\right)^L \right\} \bigg/ \left[\left(\frac{3}{4}\right)^L + 3\left(-\frac{1}{4}\right)^L\right] \nonumber\\
&=& \left\{\mathrm{Tr} \left[\prod_{j=1}^{L} {\bm{D}}^{(j)}\right] 
+ \frac{2^L}{3^L} \right\} \bigg/ \left[1 -\left(-\frac{1}{3}\right)^{L-1}\right], 
\end{eqnarray}
where
\begin{eqnarray}
\tilde{\bm{D}}^{(j)} = 
\left(\begin{array}{cccc}
\frac{1}{4} & \frac{1}{2} e^{i \theta^{(j)}} \\
\frac{1}{2} e^{-i \theta^{(j)}} & \frac{1}{4}
\end{array}\right), \ \ \ 
\tilde{\bm{E}}^{(j)} = 
\left(\begin{array}{cccc}
\frac{1}{4} & 0 \\
0 & \frac{1}{4}
\end{array}\right), 
\end{eqnarray}
and
\begin{eqnarray}
\bm{D}^{(j)} = \frac{1}{3}
\left(\begin{array}{cccc}
1 & 2 e^{i \theta^{(j)}} \\
2 e^{-i \theta^{(j)}} & 1
\end{array}\right).
\end{eqnarray}
When $q=2$, the trace of product of $\bm{D}^{(j)}$ is the leading term, then
\begin{eqnarray}
\mathrm{Tr} \left[\prod_{j=1}^{L} \bm{D}^{(j)}\right]&=&\mathrm{Tr} \left[ \frac{1}{3}
\left(\begin{array}{cccc}
1 & 2 \\
2& 1
\end{array}\right)\exp\left(i\sigma_z\frac{2\pi}{L}\right)\right]^L\nonumber\\
&=&\frac{1}{3^L}\sum_{\eta=\pm}\left(\cos\frac{2\pi}{L}+\eta\sqrt{4-\sin^2\frac{2\pi}{L}}\right)^L\nonumber\\
&=&1-\frac{\pi^2}{L}+\mathscr{O}(1/L^2).
\end{eqnarray}

\subsection{OBCs results}
Under OBCs,
AKLT model has 4-fold degenerate ground states~\cite{Affleck:1987ab,Affleck:1988ab,Fannes:1992aa}:
\begin{eqnarray}
\langle\bm{\sigma}|\Psi^{(i,j)}_{\text{AKLT}}\rangle&=&\sum_{ \alpha_1, \alpha_2, \cdots, \alpha_L\in\{1, 2\}} 
	A_{i, \alpha_2}^{\sigma_1} A_{\alpha_2, \alpha_3}^{\sigma_2} \cdots A_{\alpha_L, j}^{\sigma_L} \\
	&=& (\bm{A}^{\sigma_1}\bm{A}^{\sigma_2}\cdots\bm{A}^{\sigma_L})_{i,j},
\end{eqnarray}
where $i,j\in\{1,2\}$.
These state has spin-$1/2$ magnetization $(-1)^{i+1}/2$ and $(-1)^{j+1}/2$ around sites $1$ and $L$,
respectively.

Their norms can be calculated as
\begin{eqnarray}
\langle\Psi^{(i,j)}_{\text{AKLT}}|\Psi^{(i,j)}_{\text{AKLT}}\rangle = [\tilde{\bm{A}}^L]_{i,j}
=\frac{1}{2\times4^L} \left(\begin{array}{cc}
3^L+(-1)^L & 3^L-(-1)^L \\
3^L-(-1)^L& 3^L+(-1)^L
\end{array}\right)_{i,j}, 
\end{eqnarray}
and then,
for a fixed ground state labelled by $(i,j)\in\{1,2\}$:
\begin{eqnarray}
\langle\Psi^{(i,j)}_{\text{AKLT}}|\hat{U}^2|\Psi^{(i,j)}_{\text{AKLT}}\rangle &=& \left[\prod_{k=1}^{L} \tilde{\bm{C}}^{(k)}\right]_{i,j} \bigg/ \langle\Psi^{(i,j)}_{\text{AKLT}}|\Psi^{(i,j)}_{\text{AKLT}}\rangle\nonumber\\
&=& \frac{1}{4^L}\left[
\left(\begin{array}{cccc}
1 & 2 \\
2& 1
\end{array}\right)\exp\left(i\sigma_z\frac{2\pi}{L}\right)\right]^L{}_{i,j}\bigg/ \langle\Psi^{(i,j)}_{\text{AKLT}}|\Psi^{(i,j)}_{\text{AKLT}}\rangle\nonumber\\
&=&\left(\begin{array}{cc}1-\frac{\pi^2}{L}+i\frac{\pi}{L}&1-\frac{\pi^2}{L}-i\frac{2\pi}{L}\\
1-\frac{\pi^2}{L}+i\frac{2\pi}{L}&1-\frac{\pi^2}{L}-i\frac{\pi}{L}\end{array}\right)_{i,j}+\mathcal{O}(1/L^2)\nonumber\\
&=&1+\mathcal{O}(1/L),
\end{eqnarray}
where we can see that the asymptotical rule $\mathcal{O}(1/L)$ is still preserved under OBCs although the coefficients are modified and not the same for a general linear combination.

Similarly as Eq.~(\ref{linear_comb}) and consistently with the global $\pi$-rotation around $x$-axis:
\begin{eqnarray}\label{linear_comb}
\left.\begin{array}{c}\left[\frac{1}{\sqrt{2}}\left(\langle\Psi_\text{AKLT}^{(1,2)}|\pm\langle\Psi_\text{AKLT}^{(2,1)}|\right)\right]\hat{U}^2\left[\frac{1}{\sqrt{2}}\left(|\Psi_\text{AKLT}^{(1,2)}\rangle\pm|\Psi_\text{AKLT}^{(2,1)}\rangle\right)\right]\\\left[\frac{1}{\sqrt{2}}\left(\langle\Psi_\text{AKLT}^{(1,1)}|\pm\langle\Psi_\text{AKLT}^{(2,2)}|\right)\right]\hat{U}^2\left[\frac{1}{\sqrt{2}}\left(|\Psi_\text{AKLT}^{(1,1)}\rangle\pm|\Psi_\text{AKLT}^{(2,2)}\rangle\right)\right]\end{array}\right\}&=&1-\frac{\pi^2}{L}+\mathcal{O}(1/L^2)\in\mathbb{R}.
\end{eqnarray}

\end{document}